\documentclass[preprint,3p,twocolumn]{elsarticle}
\usepackage{graphicx}
\usepackage{multirow}

\usepackage{amssymb}
\usepackage{amsmath}

\newcommand{\bi}[1]{\bibitem{#1}}
\newcommand{\lab}[1]{\label{#1}}
\newcommand{\ba}{\begin{eqnarray}}
\newcommand{\ea}{\end{eqnarray}}
\newcommand{\be}{\begin{equation}}
\newcommand{\ee}{\end{equation}}

\newcommand{\beqs}{\begin{eqnarray}}
\newcommand{\eeqs}{\end{eqnarray}}

\newcommand{\dd}{\Delta}

\begin{document}

\begin{frontmatter}

\title{New feature in the differential cross sections at $13$ TeV  measured at the LHC}

\author{  O.V. Selyugin\footnote{selugin@theor.jinr.ru} } 
\address{ BLTP,
Joint Institute for Nuclear Research,
141980 Dubna, Moscow region, Russia }

\begin{abstract}
  Analysis of  $d\sigma/dt$ of the TOTEM Collaboration data, carried out
  without model assumptions, showed the existence of a new effect in the behavior of the
    hadron scattering amplitude at  a small momentum transfer at
    a high confidence level. A quantitative description of the data in the framework of the HEGS
    model supports the existence of  such phenomenon which can be connected with quark potentials at large
    distances.
\end{abstract}

\begin{keyword}
Hadron, elastic scattering, high energies
\end{keyword}
\end{frontmatter}

\section{Introduction}
Research into the structure of the elastic hadron scattering amplitude
   at superhigh energies and small momentum transfer - $t$
  can give a connection   between
    experimental knowledge and  the basic asymptotic theorems,
   which are based on  first principles \cite{Fisher,Block-85}.
   It gives   information about  hadron interaction
   at large distances where the perturbative QCD does not work,
   and a new theory, as, for example, instantons or string theories,
    must be developed.
There were
   many works in which the consequences of breaking the
   Pomeranchuk theorem were investigated \cite{royt}.
   It was shown  \cite{akm} that
   if the Pomeranchuk
   theorem was broken and the scattering amplitude
   grew to a maximal possible extent but did not break the Froissart boundary,
   many zeros in the scattering amplitude
   should be available in the nearest domain of
   $t \rightarrow 0$ in the limit $s \rightarrow \infty$.
   Hence, with increasing energy of colliding beams, 
   some new effects \cite{CS-PRL}
   in  differential cross sections  can be discovered at small $t$ \cite{L-range}.
 This  possibility was
 explored in some works
   \cite{barsh,kontr}.  They provided an explanation of some
   individual effects in a definite domain
   of momentum transfer
    and with a period proportional to $\Delta t$.
  In the previous paper \cite{gnsosc}, it was shown that
  "AKM oscillations" could exist
  in high-precision experimental data of the UA4/4
  Collaboration. Now we examined this effect assuming the existence
  of the potential of  hadron-hadron interactions at large distance
  and made a new  quantitative treatment of experimental data at $13$ TeV
  at a high confidence level.

\section{Hadron potential at large distance}

  The standard fast falling potentials of the Gaussian type lead
   to the exponential dependence of the scattering amplitude
   in the range of small momentum transfer.
    During a long time there have been  different attempts to find
    unusual behavior of the amplitude of the elastic hadron
    scattering. In \cite{zar}, it was shown that peripheral
   contributions of the inelastic diffraction processes led
   to the appearance in  the elastic cross sections of
   large and small periodical structures over $t$.
 It was shown in \cite{kus} that the potential
 of the rigid string led to  oscillations of the pion elastic
 form factors at large distances ($q= 10 - 20 fm^{-1}$).
  So let us take an additional potential in our case in the impact parameter space
 in the following form:
$$\chi_{osc}=h_{osc} [\omega^2 - (b+\varphi(s))^2]^{-1/2},
          \ \   b+\varphi(s) \le \omega $$  
   $$ \chi_{osc} = 0,   \ \  b+\varphi(s) > \omega, $$
here $\omega$ is the frequency and $\varphi(s)$ is the phase of
 the additional potential. This  potential leads to the scattering
 amplitude in the form
$$ F_N(s,t) \sim h_{osc} \ sin(\omega q + \varphi(s))/(\omega q + \varphi(s)) $$.

In the case of  cutting the potential, for example by
  the $\theta$-function, we
shall get a strongly oscillating expression as $t \rightarrow 0 $
 with the period, which is approximately proportional $t$, and
 with the amplitude, which is suppressed as $t$ grows.
 The potentials with the peculiarities at  the determined
distance, considered in \cite{filip},
 \ba
   g V(r) = g [1-exp( \mu (r-r_{0}))]^{-1} , \lab{vfil}
 \ea
 lead to   similar behavior of the additional term.
 In contrast with the potential of the multi-gluon exchange  there is no
  necessity to enter an additional cut  of the potential.
 But on the other hand, we have to get  eikonalization of the Born terms.
 At first, we need  to calculate the eikonal phase,
 which can be obtained directly from the interaction potential,
 and then   the amplitude in the $t$-representation
  together with the leading eikonal
  which can be directly calculated from the interaction potential.
 However, as the  potential  gives only a small additive contribution
 to the leading eikonal, as a final result we shall get
 the additional term, practically appropriate to the born term.
 Using the standard Fourier transform \cite{Gold} of the potential $V(\vec{r})$,
 one can  obtain the Born term of the scattering amplitude
 \ba
 F_{B}(t) = g \int V(\vec{r}) e^{iqr} \ d^{3}r.
  \ea
 If the potential has the spherical symmetry, the Born amplitude can be
 calculate as
  \ba
 F_{B}(t) = g/q \int_{0}^{\infty} r sin( q r )  V(\vec{r})  \ dr.
  \ea
 Numerical integration allows one to calculate an additional term
 in both  the direct and
 eikonal approach. The results of calculations are shown in
 Fig.1.
\begin{figure}
%
\begin{flushright}\vspace{-1.5cm}
 \includegraphics[width=0.45\textwidth]{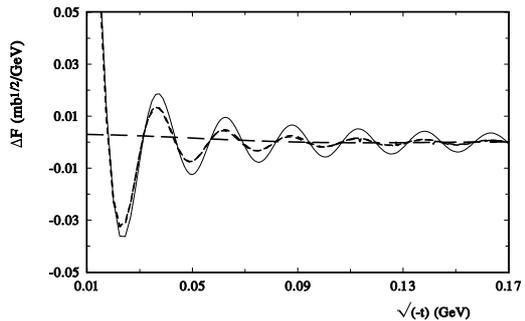}
\end{flushright}
\caption{The Born term of the amplitude, appropriate to expression (\ref{vfil}),
calculated at different $r_{0} $ (long-dashed line with $r_{0} = r_{0}^{a} $
and hard line for $r_0 = 5 r_{0}^{a} $ with $\mu = 1 \  GeV$; dashed line -
  for $r_0 = 5 r_{0}^{a} $ with $\mu = 0.1 \ GeV$
  }
\label{Fig_1}
\end{figure}

\section{Analysis of the TOTEM  13 TeV data by the selection method}

 The usual method of minimization $\chi^2$ in this situation often works
 poorly. On the one hand, we should define a certain model for  part of
 the scattering amplitude  having zeros in the domain of small $t$.
 However, this
 model may slightly differ from a real physical picture.
 On the other hand, the
 effect is rather small and gives an insignificant change
 in the sum of $\chi^2$.
 Therefore, in this work let us apply another method, namely, the method of
 comparison of two selected, statistically independent, extractions,
  for example \cite{hud}.
   If we have two statistically independent selections
  $x^{'}_{n_1}$  and $x^{"}_{n_2}$
  of values of the quantity  $X$  distributed around
  a definite value of $A$ with the standard error  equal to $1$,
  we can try
  to find the difference between
  $x^{'}_{n_1}$  and $x^{"}_{n_2}$.
  For that we can compare
  the arithmetic mean of these selections
$ \dd X = (x^{'}_1 + x^{'}_2 + ... x^{'}_{n1})/n_1  -
        (x^{"}_1 + x^{"}_2 + ... x^{"}_{n2})/n_2  =
     \overline{x^{'}_{n_1}} - \overline{x^{"}_{n_2}}.   $
  The standard deviation for that case will be
$   \delta_{\overline{x}} = [1/n_1 +1/n_2]^{1/2} $
 And if $\dd X / \delta_{\overline{x}}$ is larger than $3$, we can say
 that the difference between these two selections has  the
  $99\%$ probability .

\begin{figure}[b]
\begin{flushright}
\vspace{-1.5cm}
 \includegraphics[width=0.45\textwidth]{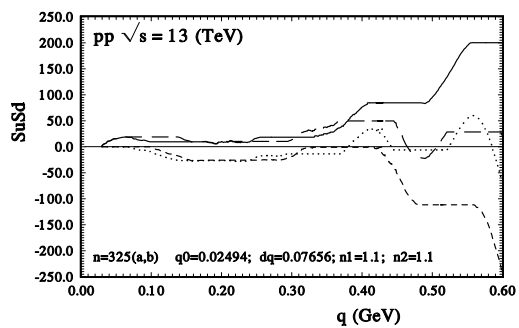}
\end{flushright}%
%
\caption{a) [left]
Sums $S^{up}$ and $S^{dn}$ calculated with additional normalization $n_i=1.1$
 for $q_{0}=0.039$ GeV and  $\delta q =0.000831$ GeV (full and dashed lines);
 and  for   $q_{0} - \delta q /2 $ ( long-dashed and dots lines).
  }
\label{Fig_2}
\end{figure}

 The deviations $\Delta R_i$ of experimental data from these
 theoretical cross
 sections will be measured in units of experimental error for
 an appropriate point
\ba
\Delta R_ {i} = [(d\sigma/dt_{i})^{exp}
  -  (d\sigma/dt_{i})^{th}] / \delta_{i}^{exp},
\ea
 where $\delta_{i}^{exp} $ is an experimental error.
 To take this effect into account, we  break
 the whole studied interval of momentum transfer into $k$  equal pieces
 $k \delta t = (t_{max} -t_{min}) $, where
   $\Delta R_{i} $ is summed
 separately over even and odd pieces.
 Thus, we get two sums $S^{up} $ and $S^{dn} $ for
  $n_1$ even and $n_2$ odd interval. At this
  $n_1 + n_2 = k$ and $|n_1 - n_2| = 0$ or $1$
  \ba
   S^{up} &=& \sum_ {j=1}^{n_1} (\sum^{N}_{i}
            \Delta R_{i})|_{\delta q (2j-1) < q_i \leq \delta q(2j) },  \nonumber \\
S^{dn} &=& \sum_{j=1}^{n_2} (\sum^{N}_{i=1} \Delta R_{i})
|_{\delta q (2j) < q_i \leq  \delta q (2j + 1) }.
\label{sdn}
\ea
 In the case of some difference of experimental data from
 the theoretical behavior
 expected by us or incorrectly determined parameters, these two sums
 will deviate from zero; however, their sizes should coincide within
 experimental errors.
 However, this will be so in the case if experimental data have no any
 periodic structure or a sharp effect coincides with one interval.
 We assume that such a periodic structure is available
 and its period coincides with the chosen interval $2 \delta t$. In this
  case, the sum
 $S^{up} $ will contain, say, all positive half-cycles;
 and the sum $S^{dn} $, all negative half-cycles.
 The difference between $S^{up} $ and
 $S^{dn}$ will show the magnitude of an additional effect summed
 over the whole
 researched domain.

 Our method  does not require  exact representation of the
 oscillatory part of the scattering amplitude, and now let us apply it to  new LHC data
 of the TOTEM Collaboration at $13$ TeV  \cite{TOTEM13-1set,TOTEM13-2set}.
 It gives us two sets of the data: one  $0.000879 <|t|< 0.201041$ GeV$^2$ includes the Coulomb-hadron
 interference range and the other $0.0384 <|t|< 3.822$ GeV$^2$  for the large $t$.
 Both sets have the overlapping region.  We find that three first points of the second set
 hardly differ from the data of the first set and we removed them.
For the first analysis we do not include the region of the diffraction minimum
 as we try, using this new method, to examine some
additional oscillation behaviour with minimum model assumption,
 but the region of the dip requires some model
for the description of the elastic scattering in a wide region of $t$,

Here and below we use only statistical errors, and systematic errors are taken into account
as an additional normalization coefficient for both the sets.
For the basic (non-oscillating ) amplitude
we use the standard exponential form with three slopes multiplied by $t, t^2, t^3 $.
In Fig.2, such sums in eq.(3) are represented for the period that is proportional to $\sqrt{-t}$.
Then we move the  segments by one-half the segment and calculate
these sums again. The results are shown in Fig.2 by the long-dashed and dotted lines.
The large difference between the first and second cases clearly shows the existence of some oscillation contributions in the scattering amplitude especially at large $t$.
Now let us calculate such sums with the period that is proportional to $t$.
The results are shown in Fig.3 a,b.
 To evaluate the size of a possible effect, one should examine the difference
 of the arithmetic mean values
   $\dd S$ and the corresponding dispersion - $\delta S$ \cite{hud}
\ba
\overline{ \dd S} = \overline{S^{up}} - \overline{S^{dn}}; \ \ \overline \delta S = (1/[1/n_1 + 1/n_2]^{1/2})/N.
\ea

\begin{figure}[t]
\begin{flushleft}
\vspace{-1.5cm}
 \includegraphics[width=0.45\textwidth]{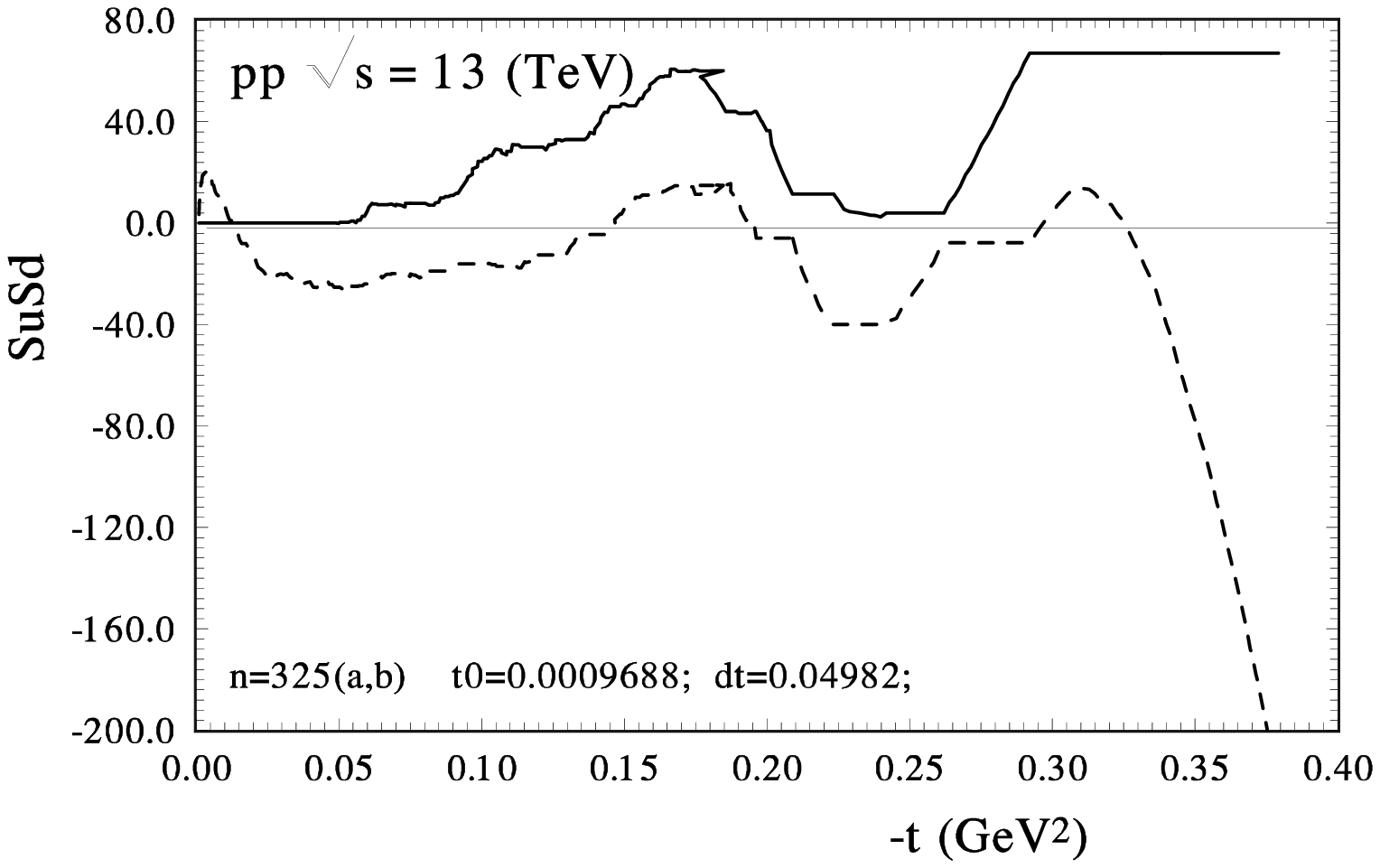}
 \end{flushleft}%
\begin{flushleft}
\vspace{-1.5cm}
  \includegraphics[width=0.45\textwidth]{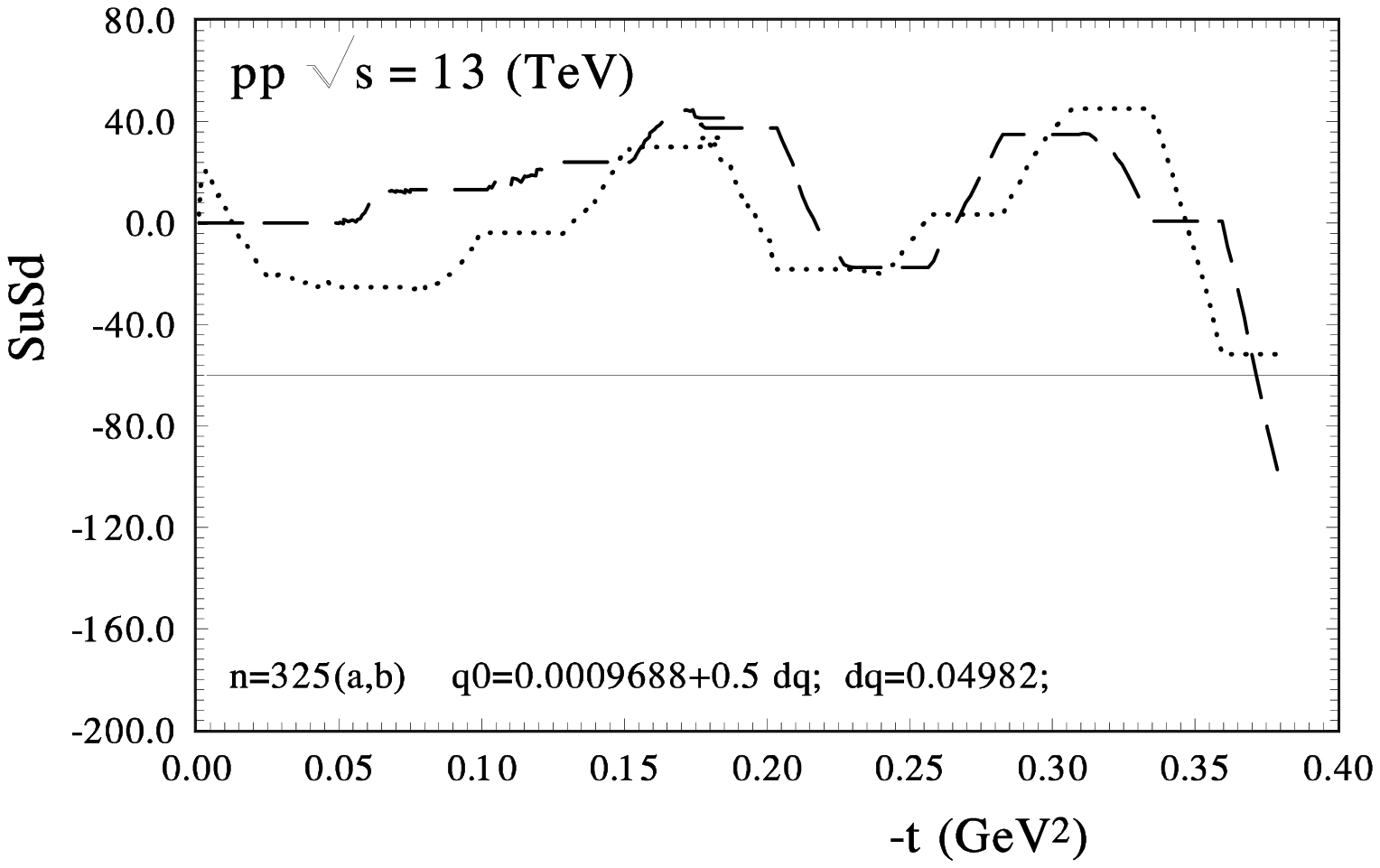}
\end{flushleft}
\caption{
 $S^{up}$ and $S^{dn}$ calculated with additional normalization $n_i=1.1$
 for $t_{0}=9 \dot 10^{-3}$ GeV$^2$ and  $\delta t =0.0498$ GeV$^2$ (full and dashed lines);
 b) [bottom] the same for   $q_{0} - \delta q /2 $ ( long-dashed and dots lines).
  }
\label{Fig_3}
\end{figure}

\begin{figure}[t]
\begin{flushleft}
\vspace{-1.5cm}
 \includegraphics[width=0.4\textwidth]{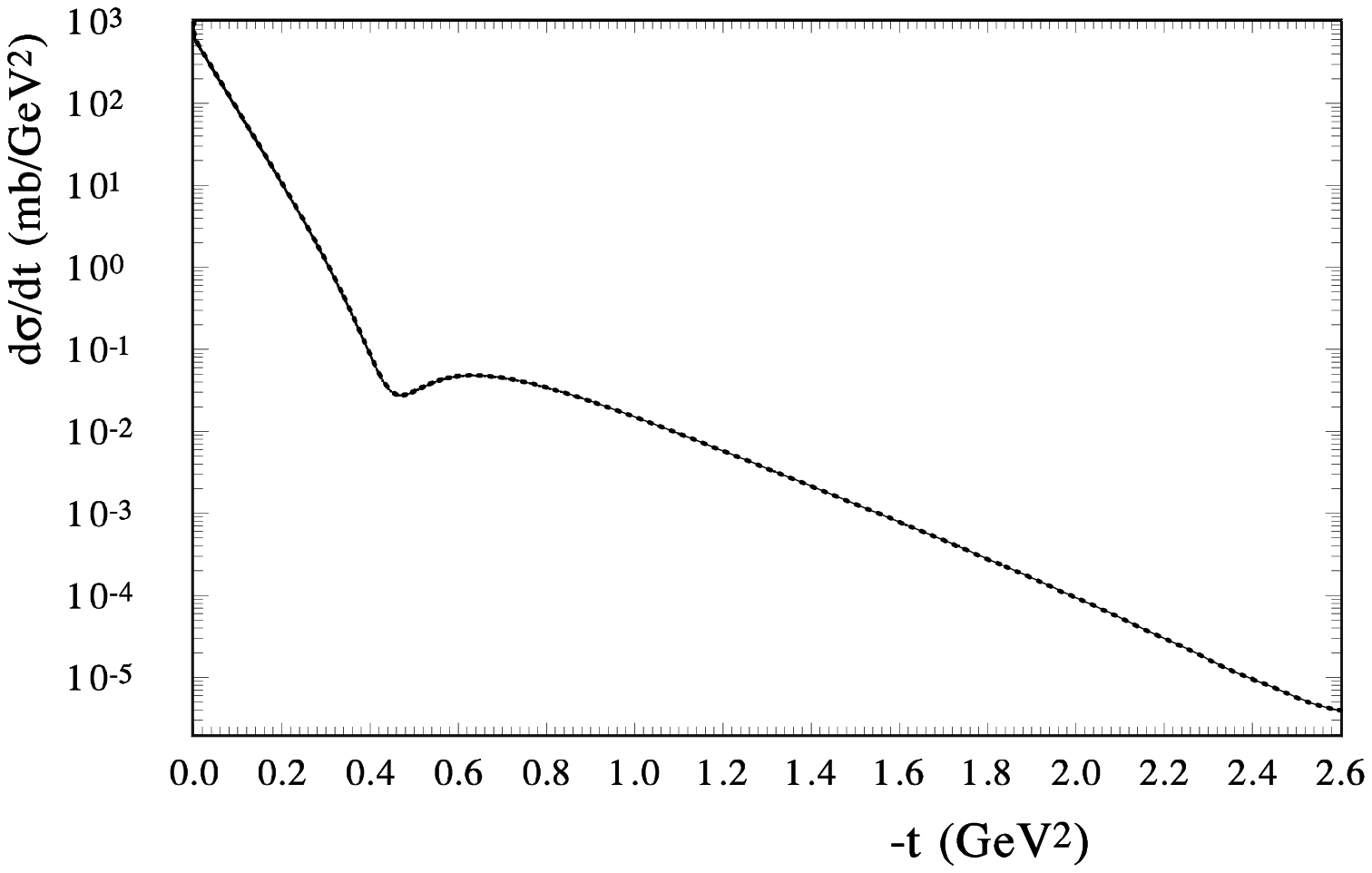}
\end{flushleft}%
\begin{flushleft}
\vspace{-1.cm}
  \includegraphics[width=0.4\textwidth]{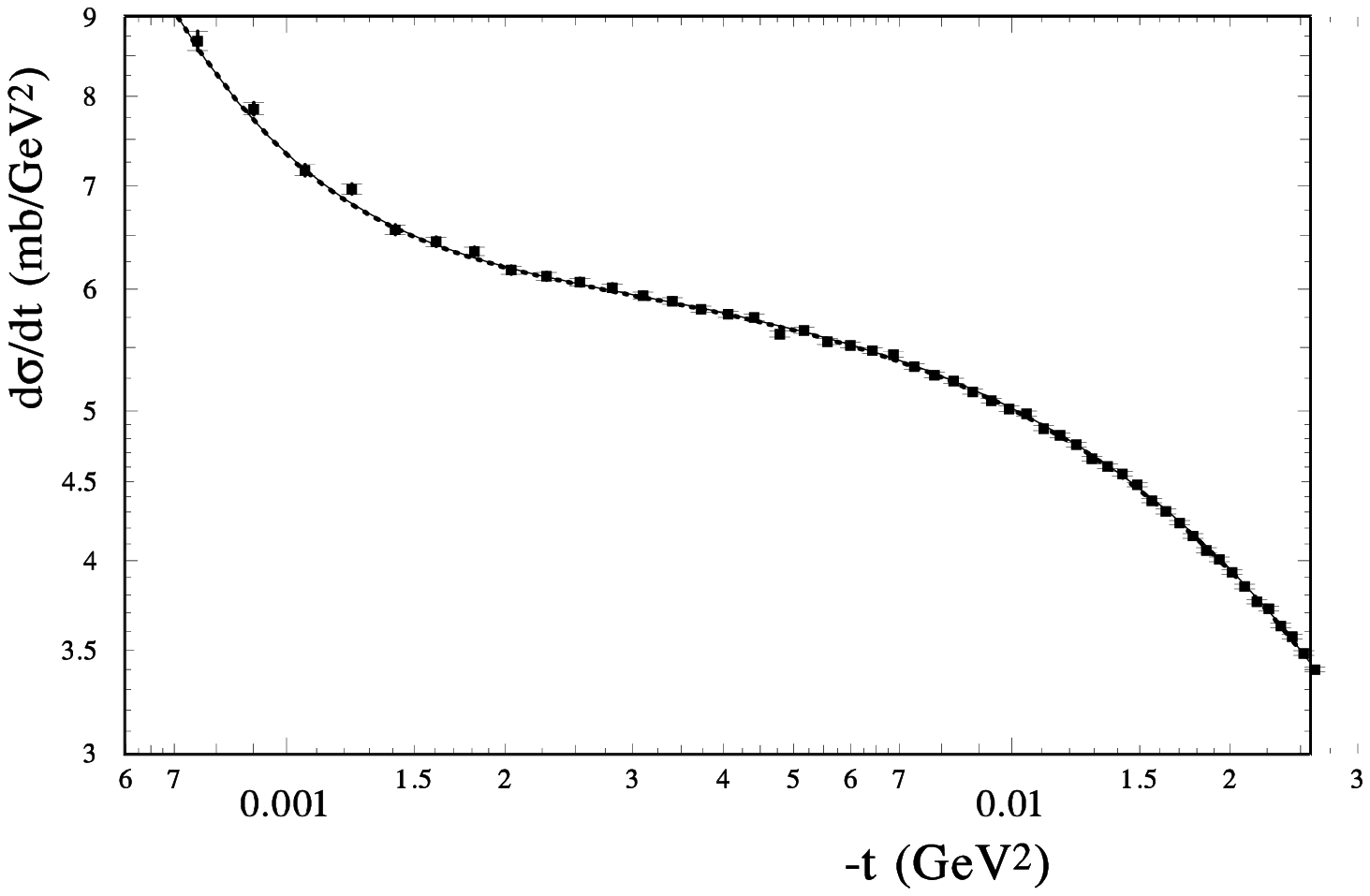}
\end{flushleft}
\begin{flushleft}
\vspace{-1.cm}
   \includegraphics[width=0.4\textwidth]{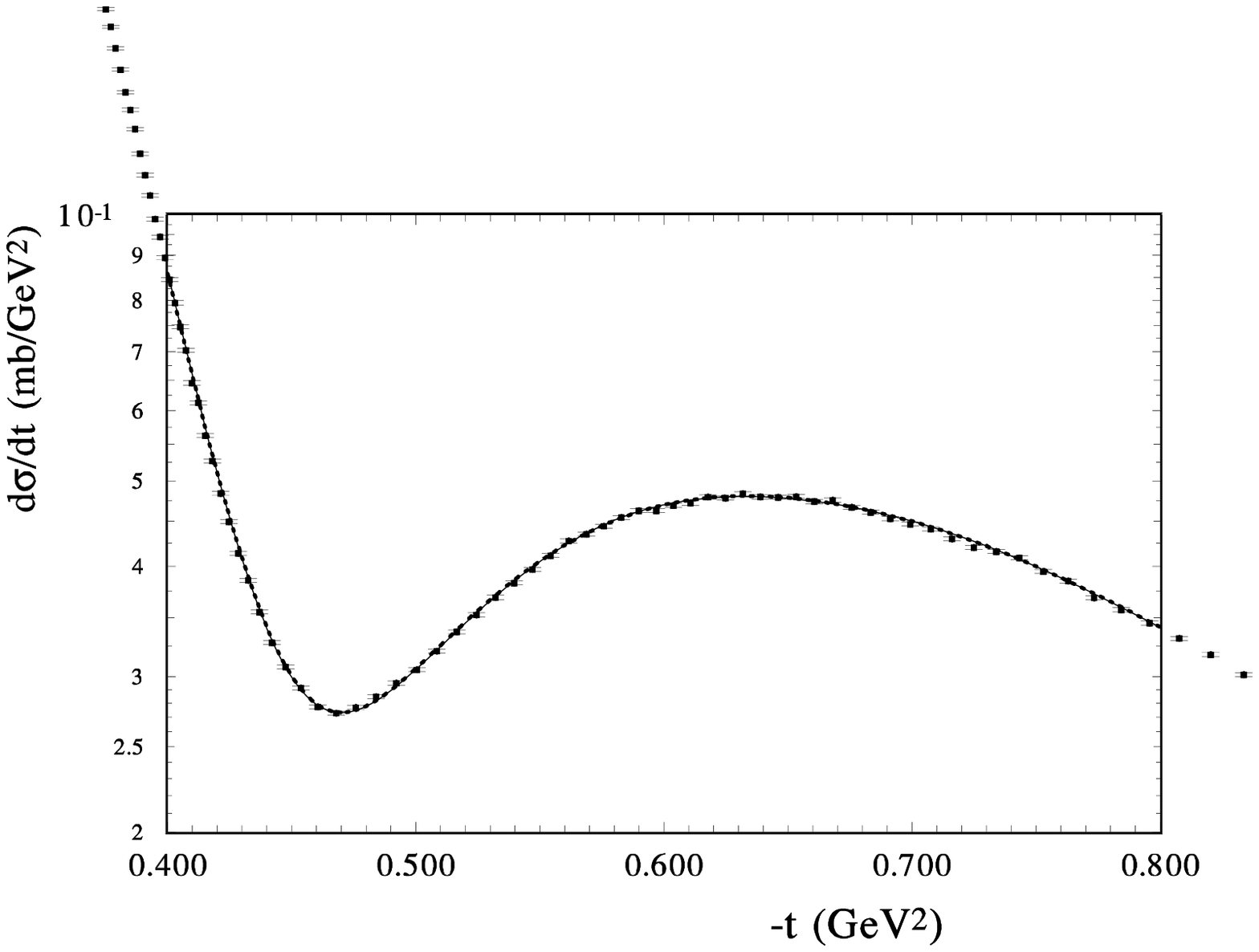}
\end{flushleft}
\caption{a) (top) The $d\sigma/dt$ of the elastic $pp$ scattering at
 $13 $ TeV (the data of the two sets of the TOTEM Callaboration -small points;
 the thin line  -the model calculations without the oscillation  contribution;
   the tiny-dashed thick line - with the oscillation contributions;
   b) (midle) the same  - magnification at small $t$;
   c) (bottom) the same - magnification of the region of the diffraction minimum
  }
\label{Fig_4}
\end{figure}

  Let us calculate the sum of $\Delta R_{i} $ and its
 arithmetic mean
\ba
\overline {\Delta S} =285/325 = 0.877 \pm 0.028.
\ea
  Obviously, the existence of the oscillatory contributions at a high
  confidence level  is show.
%

\section{HEGS model analysis}

 Now let us try to find the form of such additional oscillation contribution
   to the basic elastic scattering amplitude.
   As a basis, take our high energy generalized structure (HEGS) model \cite{HEGS0,HEGS1} which quantitatively  describes, with only a few parameters, the   differential cross section of $pp$ and $p\bar{p}$
  from $\sqrt{s} =9 $ GeV up to $13$ TeV, including the Coulomb-hadron interference region and the high-$|t|$ region  up to $|t|=15$ GeV$^2$,
  and quantitatively well describes the energy dependence of the form of the diffraction minimum \cite{HEGS-min}.
   However, to avoid  possible problems
 connected with the low-energy region, we consider here only the asymptotic variant of the model \cite{HEGSh}.
   The total elastic amplitude in general receives five helicity  contributions, but at
   high energy it is enough to write it as $F(s,t) =
  F^{h}(s,t)+F^{\rm em}(s,t) e^{\varphi(s,t)} $\,, where
 $F^{h}(s,t) $ comes from the strong interactions,
 $F^{\rm em}(s,t) $ from the electromagnetic interactions and
 $\varphi(s,t) $
 is the interference phase factor between the electromagnetic and strong
 interactions \cite{Bethe,Can,Petr,selmp1}.
 Note, that in [23] insist that the Bethe form for the Coulomb-nuclear scattering amplitude and the same amplitude based
 on the additive eikonal are incompatible.
    The Born term of the elastic hadron amplitude at large energy can be written as
    a sum of two pomeron and  odderon contributions,
    \begin{eqnarray}
 F_{\mathbb{P}}(s,t)=\hat{s}^{\epsilon_0}\left(C_{\mathbb{P}} F_1^2(t)  \ \hat s^{\alpha' \ t} + C'_{\mathbb{P}}  A^2(t) \ \hat s^{\alpha' t\over 4} \right) \; , 
 \end{eqnarray} 
    \begin{eqnarray}
 F_{\mathbb{O}}(s,t)=i \hat{s}^{\epsilon_0+{\alpha' t\over 4}} \left( C_{\mathbb{O} }
   + C'_{\mathbb{O}} t/(1-r_{\mathbb{O} }^{2} t ) \right) A^2(t).
 \end{eqnarray}
 All terms are supposed to have the same intercept  $\alpha_0=1+\epsilon_0 = 1.11$, and the pomeron
 slope is fixed at $\alpha'= 0.24$ GeV$^{-2}$.
  The model takes into account two hadron form factors $F_1(t)$ and $A(t)$, which correspond to  the charge and matter
  distributions \cite{GPD-PRD14}. Both the form factors are calculated  as the first and second moments of  the same Generalized Parton Distributions (GPDs).
 As a probe for  the oscillatory function take
\vspace{-0.1 cm}
\ba
 f_{osc}(t)=h_{osc} (i+\rho_{osc}) J_{1}(\tau))/\tau; \nonumber  \\
  \tau = \pi \ (\phi_{0}-t)/t_{0},
\ea
here $J_{1}(\tau)$ is the Bessel function of the first order.
 This form has only a few additional fitting parameters and allows one to represent
 a wide range of  possible oscillation functions.

  After the fitting procedure,
  with the modern version of FUMILY \cite{Sitnik},
    we obtain  $\chi^2/dof =1.24$ (remember that we used only statistical errors).
 One should note that the last points of the second set above $-t=2.8$ GeV$^2$  show
 an essentially different slope, and we removed them. The total number of  experimental points
 of both sets equals $415$. If we remove the oscillatory function, then
  $\chi^2/dof =2.7$, so an increase is more than two times.  If we make a new fit without $f_{osc}$,
 then  $\chi^2/dof =2.5$ decreases but remains large. Our model calculations are represented
  in Fig.4(a,b,c). It can be seen that the model gives a beautiful description of the differential
  cross section in both  the Coulomb-hadron interference region and  the region of the diffraction dip.

  To see the oscillations in the differential cross sections, let us
  determine two values - one is pure theoretical and the  other with experimental data

\ba
 R\Delta_{th} &=& \frac{[d\sigma/dt_{th0+osc} - d\sigma/dt_{th0}]}{d\sigma/dt_{th0}}, \nonumber \\
 R\Delta_{Exp}&=& \frac{[d\sigma/dt_{Exp} - d\sigma/dt_{th0}]}{d\sigma/dt_{th0}}.
\ea
  The corresponding values calculated from the fit of two sets of the TOTEM data at $13$ TeV are presented in Fig.5.
  At small $t$ there is a large noise;  however, the oscillation
  contributions can be seen. This corresponds to the small size of the $SuSd$ values in
   Fig.3; however, at large $-t> 0.1$ GeV$^2$ we can see that $R\Delta_{th}$
    is similar to the value $R\Delta_{Exp}$. The oscillation contribution is small;
     however, the noise of
    the background  decreases at this $t$ and  does not dump
    the oscillation part.
\begin{figure}[t]
\begin{flushright}
\vspace{-1.5cm}
 \includegraphics[width=0.45\textwidth]{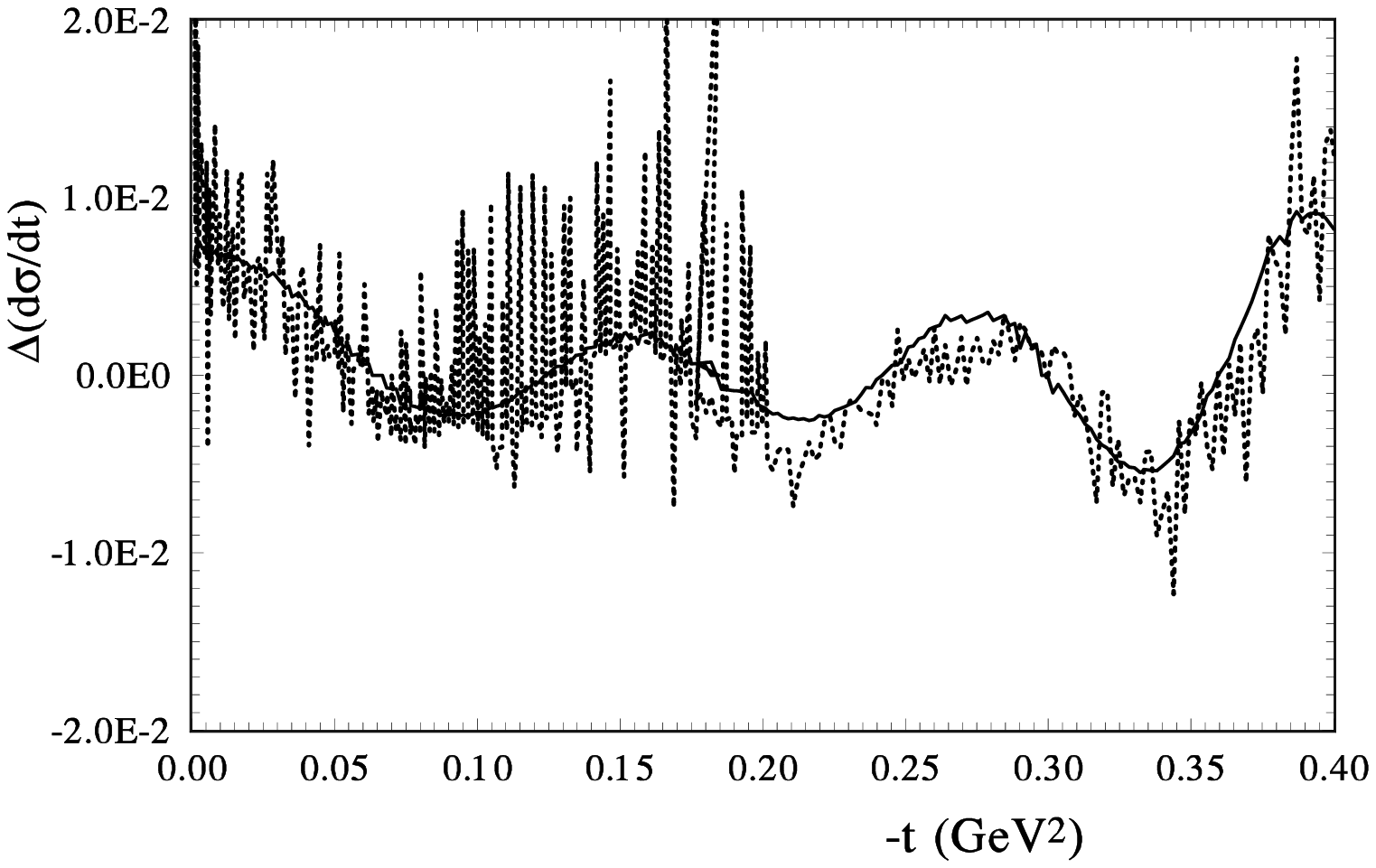}
\end{flushright}
\begin{flushright}
\vspace{-1.cm}
  \includegraphics[width=0.45\textwidth]{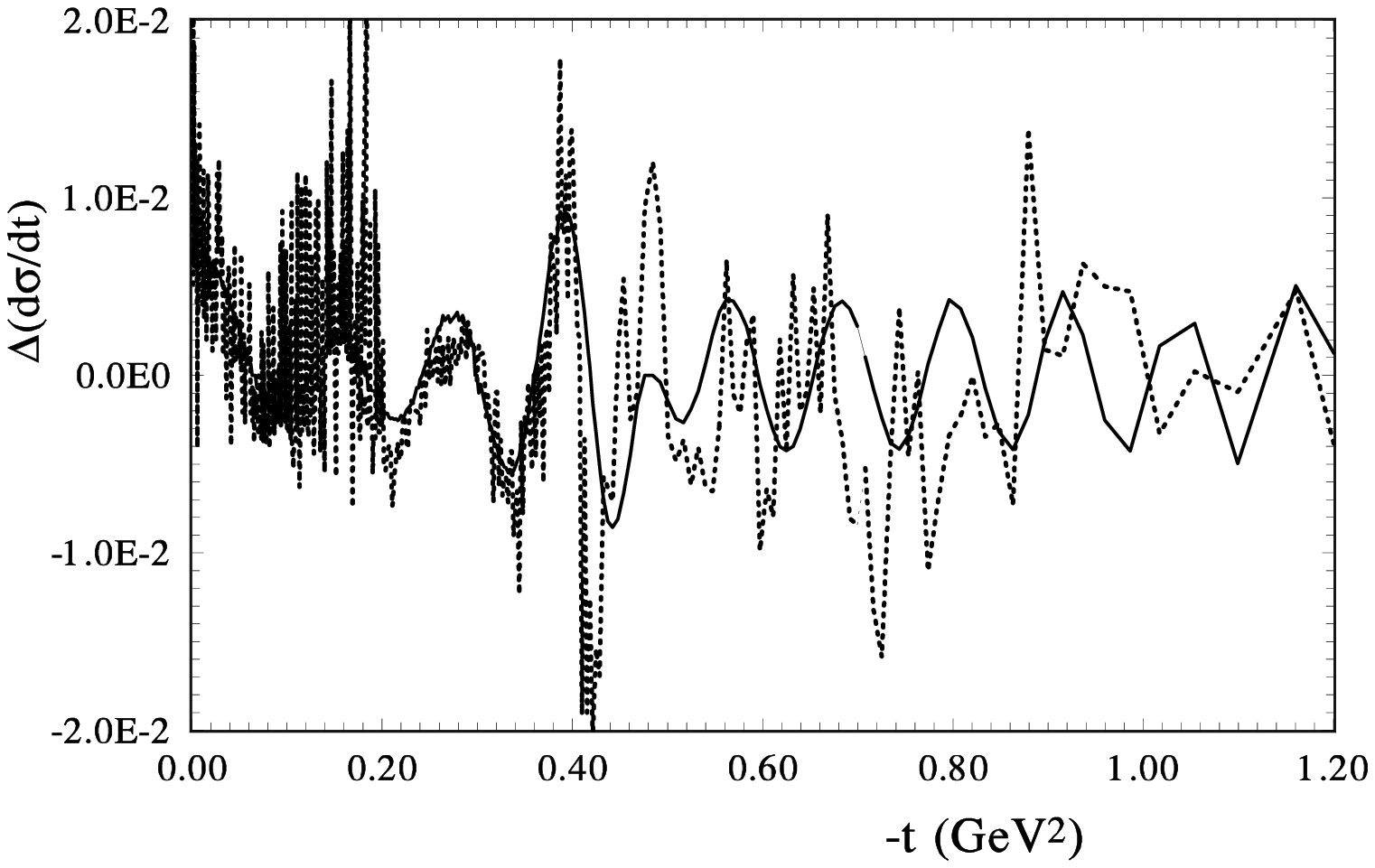}
\end{flushright}
\caption{ $R\Delta{th} $ of eq.(9a) (the hard line) and $R\Delta{ex} $
  eq.(9b) (the tiny line) [top - region of small $t$; bottom - large $t$ ]
 at  $\sqrt(s)=13 $ TeV.
  }
\label{Fig_5}
\end{figure}
\section{Conclusion}

   We have shown that HEGS model describes at a quantitative level the new experimental data at $13$ TeV
      with taking into account only statistical errors.
      However, 
       only adding  the small oscillatory term allows one to  obtain $\chi^{2}_{dof}=1.25$.
      The phenomenon of  oscillations  of the elastic scattering amplitude
  will give us important information about
      the behavior of the hadron interaction potential at large distances.
      We have shown the existence of such oscillations
      at the statistical  level by three methods: a) the method of  statistically independent selection;
      b) the comparison of the $\chi^2$ without oscillation ($\sum \chi^2 =1140$) and with
      oscillation ($\sum \chi^2 = 515$); c) the comparison of $R\Delta_{th}$ and $R\Delta_{exp}$, Fig.5).
      All three methods show the presence of the oscilatory behavior.
      Very likely that such effects exist also in  experimental data
      at essentially smaller energies \cite{osc-conf}
      but, maybe, they have a more complicated form
      (with two different periods, for example).
      The phenomena of  oscillations are also related to
       the asymptotic properties of the scattering amplitude.
     They can impact  the determination of
     the sizes of the total cross sections, the ratio of
    the elastic to the total cross sections and the size of the
    $\rho(s,t)$ - the ratio of the real to imaginary part of the elastic scattering
    amplitude.

\section*{Acknowledgements}{OVS would  like to thank J.-R. Cudell for the fruitful discussion
 and  the  University of Li\`{e}ge
  where part of this work was done. This work was also  supported
by the Fonds de la Recherche Scientifique-FNRS, Belgium, under grant No. 4.4501.15
}

\end{document}